\documentclass[10pt,twocolumn,letterpaper]{article}

\usepackage{wacv}              

\usepackage{graphicx}
\usepackage{amsmath}
\usepackage{amssymb}
\usepackage{booktabs}
\usepackage{multirow}
\usepackage{multicol}

\usepackage[pagebackref,breaklinks,colorlinks]{hyperref}

\usepackage[capitalize]{cleveref}
\crefname{section}{Sec.}{Secs.}
\Crefname{section}{Section}{Sections}
\Crefname{table}{Table}{Tables}
\crefname{table}{Tab.}{Tabs.}

\begin{document}

\title{Conditional GAN for Enhancing Diffusion Models in Efficient and Authentic Global Gesture Generation from Audios}

\author{
    \begin{minipage}{0.3\textwidth}
        \centering
        Yongkang Cheng \\
    \end{minipage}
    \hfill
    \begin{minipage}{0.3\textwidth}
        \centering
        Mingjiang Liang \\
    \end{minipage}
    \hfill
    \begin{minipage}{0.3\textwidth}
        \centering
        Shaoli Huang\thanks{corresponding author} \\
    \end{minipage}
    \hfill
    \\
    \\
    \begin{minipage}{0.3\textwidth}
        \centering
        Gaoge Han \\
    \end{minipage}
    \hfill
    \begin{minipage}{0.3\textwidth}
        \centering
        Jifeng Ning \\
    \end{minipage}
    \begin{minipage}{0.3\textwidth}
        \centering
        Wei Liu \\
    \end{minipage}
}

\maketitle

\begin{abstract}
   Audio-driven simultaneous gesture generation is vital for human-computer communication, AI games, and film production. While previous research has shown promise, there are still limitations. Methods based on VAEs are accompanied by issues of local jitter and global instability, whereas methods based on diffusion models are hampered by low generation efficiency. This is because the denoising process of DDPM in the latter relies on the assumption that the noise added at each step is sampled from a unimodal distribution, and the noise values are small. DDIM borrows the idea from the Euler method for solving differential equations, disrupts the Markov chain process, and increases the noise step size to reduce the number of denoising steps, thereby accelerating generation. However, simply increasing the step size during the step-by-step denoising process causes the results to gradually deviate from the original data distribution, leading to a significant drop in the quality of the generated actions and the emergence of unnatural artifacts. In this paper, we break the assumptions of DDPM and achieves breakthrough progress in denoising speed and fidelity. Specifically, we introduce a conditional GAN to capture audio control signals and implicitly match the multimodal denoising distribution between the diffusion and denoising steps within the same sampling step, aiming to sample larger noise values and apply fewer denoising steps for high-speed generation. In addition, to enable the model to generate high-fidelity global gestures and avoid artifacts, we introduce an explicit motion geometric loss to enhance the quality and global stability of the generated gestures. Numerous qualitative and quantitative experiments show that compared to contemporary diffusion-based methods, our method offers faster generation speed and higher fidelity, and compared to non-diffusion methods, it provides a more stable global effect and a more natural user experience.
\end{abstract}

\begin{figure*}
  \begin{center}
    \includegraphics[width=1\linewidth]{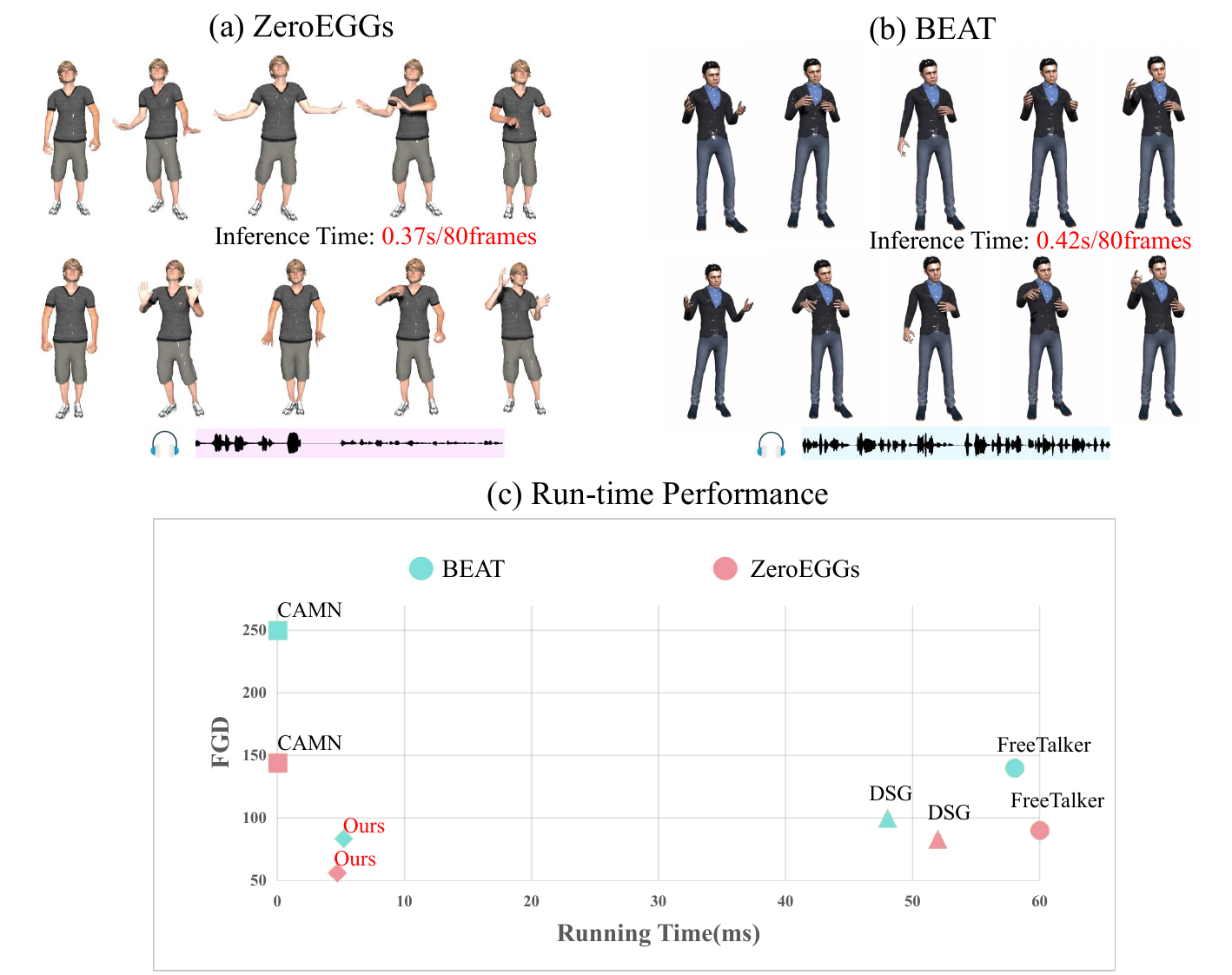}
    \vspace{-1em}
  \end{center}
  \caption{Our method generates global gesture motion highly matched with input audio melody in a short runtime. On the (a) BEAT dataset and (b) ZeroEGGs dataset, our method's average runtime is 5ms and 4.6ms per frame, respectively. For reference, DSG's corresponding times are 5.2 and 4.8s. (c) The overall comparison of inference time cost and FGD on both datasets. In the figure, we compare the per-frame runtime with the state-of-the-art methods' FGD.}
  \label{fig:title_img}
  \end{figure*}

\section{Introduction}
\label{sec:intro}
\noindent Body movements and gesture sequences play a pivotal role in human communication by conveying non-verbal cues to accurately express intent. The objective of simultaneous gesture generation is to create diverse and dynamic human postures synchronized with audio cues. Generating spontaneous co-speech gestures for non-player characters (NPCs) in virtual reality, gaming, and film can offer a more immersive user experience, a topic that has recently garnered substantial research interest. Some generation methods based on GANs and VAEs show promise but are limited by strict motion representation requirements and jittery arm posture generation, respectively. Diffusion-based generation methods, such as DiffGesture, are at the cutting edge in terms of generation quality, as they can effectively capture target distributions of various gestures. However, these methods suffer from low generation efficiency; for instance, DiffGesture requires 6 seconds to generate an 88-frame gesture sequence slice. The inability to achieve rapid generation while maintaining high gesture quality hinders their effectiveness in real-world applications, such as real-time conversational virtual humans.

With concerted efforts, strategies to mitigate the low efficiency of diffusion-based generation methods have bifurcated into two primary approaches: 1) latent space diffusion, and 2) the DPM-Solver technique. The former, initially proposed by MLD and validated in GestureClip, learns the latent space of global gestures and subsequently performs diffusion on low-dimensional, high-density latent variables. However, this two-stage method's performance is contingent on the gesture embedding space learned in the initial stage, making the learning of an effective embedding space a challenging task. The expressiveness of this space often dictates the outcome of downstream gesture generation. Conversely, the DPM-Solver approach, tailored explicitly for diffusion models, commences with the standard 1000-step DDPM denoising method. It first analyzes linear terms in diffusion steps, substitutes complex integral terms using the logarithm of the signal-to-noise ratio, and finally obtains a noise estimate via Taylor expansion of the exponential integral of the noise prediction model. Despite the accurate computation of known terms, approximation estimation of the neural network part still introduces discretization errors. Moreover, the first-order expansion form of DPM-Solver, the DDIM sampling strategy, expedites generation by increasing step size and decreasing step count. However, it overlooks that DDPM's denoising is predicated on a Gaussian distribution assumption with minuscule noise values at each step. This oversight leads to the DDIM sampling strategy skipping many reverse steps, resulting in sampled noise often emanating from a denoising distribution more intricate than the Gaussian distribution, thereby compromising the generation effect. Higher-order expansion methods, due to the presence of higher-order derivatives, necessitate multiple invocations of the denoising function, thereby prolonging the time spent per step, as demonstrated in the quantitative experiments presented in Section 4. Consequently, accurately capturing more complex multimodal distributions becomes an imperative for several step sampling methods.

Compared to DDIM's strategy of directly stacking denoising steps, we have astutely pinpointed the root of the problem. Upon reevaluating the denoising process of DiffGesture, we discern that the method hinges on the assumption that the posterior noise added at each denoising step is sampled from a unimodal distribution and is relatively small. With a sufficient number of steps, it can denoise the original motion sequence from the standard Gaussian distribution. Consequently, the inefficiency of previous methods stems from small step noise sampling and a large number of sampling steps. Motivated by this challenge, and drawing inspiration from the latest work in the fast text-to-image domain ~\cite{xu2023semi}, we introduce a framework for swiftly generating expressive speech gesture sequences. To the best of our knowledge, this is the inaugural work in the speech gesture generation domain to consider real-time performance, thereby breaking free from the constraints of prior methods and paving the way for advancements in this field. Our objective is to increase step size and diminish generation steps, essentially modeling the intricate multi-step sampling process. Unlike the fully implicit matching constraint method of DDGAN ~\cite{xie2021dual} in the image generation domain, we incorporate an explicit constraint on motion geometric loss to circumvent the introduction of detrimental motion noise during the training process.

Specifically, we guide the GAN to model the distribution by conditioning on the time step t, enabling the model to capture complex denoising distributions with a specified number of steps (or step sizes). Simultaneously, we use audio signals (style labels and spectral features) as control conditions to guide the model in capturing the distribution of heterogeneous motions. This implicit matching adversarial learning strategy employs a conditional GAN to match the conditional distribution between the diffusion and denoising processes, allowing for the addition of large random noise between adjacent diffusion steps, and achieving denoising in just a few steps. Moreover, we introduce explicit motion constraints in the form of motion geometric loss to alleviate motion jitter issues. Notably, our framework is an easy-to-train end-to-end model. 

Our method's capability to swiftly generate high-quality co-speech gesture sequences is showcased in Figure~\ref{fig:title_img}, achieving a time cost reduction of approximately 12.35 times compared to DiffuseStyleGesture. Additionally, we carried out extensive experimental assessments, substantiating that our approach considerably surpasses contemporary diffusion-based methods in terms of generation efficiency and greatly outperforms non-diffusion methods in quality.
\section{Related Work}
\noindent\textbf{Co-speech gesture generation} is an exceptionally intricate task that necessitates understanding speech melody, semantics, gesture motion, and their mutual relationships. Early data-driven approaches (e.g., those proposed in ~\cite{liu2022audio,habibie2021learning}) strive to learn gesture matching from human demonstrations but often yield limited motion diversity. Subsequent research (such as ~\cite{habibie2021learning,yi2023generating,xie2022vector}) bolsters models' capacity to generate diverse outcomes and introduces the notion of generating unique and expressive gesture results. Some studies, like ~\cite{yang2023diffusestylegesture,yang2023unifiedgesture,ahuja2020style,ao2023gesturediffuclip}, train a unified model for multiple speakers, embedding each speaker's style in the space or incorporating style transfer techniques. Other works ~\cite{zhou2022gesturemaster,habibie2022motion} employ motion matching to generate gesture sequences, but these methods often demand intricate matching rules. Despite the challenges, audio-driven animation has garnered considerable attention.

\noindent\textbf{Diffusion generation models} have garnered remarkable achievements across multiple domains \cite{rombach2022high,chen2023executing,saharia2022palette}, particularly in simultaneous gesture generation. Specifically, DiffGesture pioneers the integration of diffusion models into audio-controlled gesture generation methods, recovering gesture actions through stepwise noise prediction. DiffStyleGes is a landmark work that introduces motion diffusion models to reconstruct original motion representations by learning the relationship between motion representations and input audio control conditions, while also providing stylized hard labels for controlling the generated gesture style. Recent research, FreeTalker~\cite{yang2024freetalker} and ExpGest~\cite{cheng2024expgest}, further concentrates on global gesture generation methods under the joint control of text and audio. In our work, we leverage attention mechanisms to capture the relationship between gesture sequences and speech, generating results that closely match speech. However, due to the high dimensionality and iterative nature of diffusion models, runtime generation based on the original diffusion model DDPM has been beset by time overhead. GestureClip introduces the concept of latent diffusion into runtime generation, reducing computational resource requirements by diffusing from low-dimensional, high-density embedding spaces through latent variable sampling. Its core idea entails first training a VAE for gesture motion embedding and subsequently applying latent diffusion in the learned latent space. However, this two-stage method is a non-end-to-end training approach. Moreover, merely stacking noise steps and discarding redundant denoising steps result in artifacts that are unacceptable to humans. In contrast, our method is an end-to-end model. In this paper, we propose a conditional GAN for implicit matching constraints, modeling complex distributions (potentially multimodal distributions that deviate from the DDPM\cite{ho2020denoising} assumption) in multi-step processes through contrastive learning strategies. Simultaneously, we explicitly constrain the human body geometric loss, maintaining high fidelity of human motion while enhancing generation speed.

\begin{figure*}[t]
  \begin{center}
    \includegraphics[width=1\linewidth]{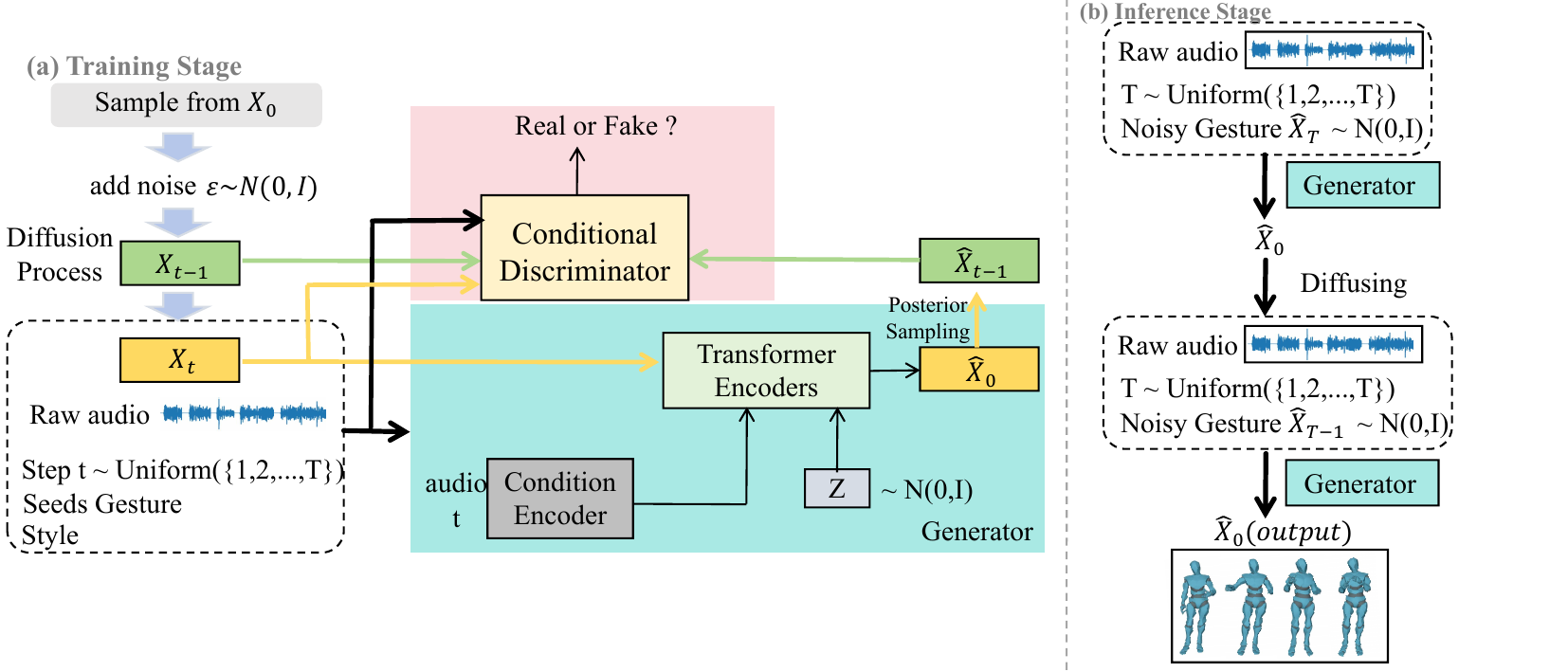}
  \end{center}
  \caption{\textbf{Network architecture.} During training, we introduce a GAN structure based on conditional denoising diffusion to capture the complex distribution of gesture sequences in a multi-step process, enabling larger sampling step sizes. During inference, we use large step sizes and fewer steps for sampling, according to the input audio control signal, to achieve fast, high-quality gesture sequences, thus supporting real-time tasks.}
  \label{fig:arc}
  \end{figure*}
\section{Method}
Our objective is to accelerate the denoising process with fewer steps and larger step sizes, enabling our model to efficiently generate high-fidelity and diverse co-speech gesture sequences based on the input audio signal in a shorter time.
\subsection{Diffusion Model for Generating Gestures}
Gesture sequences are represented as $x^{1:N}$ and are associated with corresponding conditional control signals $c$, such as audio signals~\cite{yang2023diffusestylegesture}, where $N$ denotes the number of frames in the gesture sequence. Unconditional gesture generation can also be achieved by setting the conditional signal to $\emptyset$. In this paper, we use a diffusion probabilistic model~\cite{ho2020denoising} to generate co-speech gestures, where the diffusion model gradually anneals pure Gaussian noise into a gesture distribution $p(x)$. Consequently, the model can predict noise from a T-timestep Markov noise process $\{x_{t}^{1:N}\}_{t}^{T}$, where $x_{0}^{1:N}$ is directly sampled from the original data distribution. The diffusion process is as follows:
\begin{equation}
\begin{aligned}
 q(x_{t}|x_{t-1}) = \mathcal{N}(x_{t};\sqrt{1-\beta_{t}}x_{t-1},\beta_{t}I), \\
 = \mathcal{N}(\sqrt{\frac{\alpha_{t}}{\alpha_{t-1}}}x_{t-1},(1-\frac{\alpha_{t}}{\alpha_{t-1}})I), 
\end{aligned}
\end{equation}
where $\{\beta_{t}\}_{t=1}^{T}$ is the variance schedule and $\alpha_{t} = \prod_{s=1}^{t}(1-\beta_{s})$. Then the reverse process becomes $p_{\theta}(x_{0:T}):=p(x_{T})\prod_{t=1}^{T}p_{\theta}(x_{t-1}|x_{t})$, starting from $x_{T}\sim\mathcal{N}(0,I)$ with noise predictor $\epsilon_{t}^{\theta}$:
\vspace{-5pt}
\begin{equation}
\begin{aligned}
 x_{t-1}= \frac{1}{\sqrt{1-\beta_{t}}}(x_{t}-\frac{\beta_{t}}{\sqrt{1-\alpha_{t}}}\epsilon_{t}^{\theta}(x_{t}))+\sigma_{t}z_{t},\\
\end{aligned}
\end{equation}
where $z_{t} \sim \mathcal{N}(0,I)$ and $\sigma_{t}^{2}=\beta_{t}$ means the variance schedule stays constant.\\
However, unlike the original DDPM~\cite{ho2020denoising}, we consider the extreme physical constraints inherent in 3D human bodies and deviate from image generation by replacing predicted noise with original human representations. Consequently, at each step of the denoising process, we reconstruct the original representation from pure Gaussian noise, ultimately producing the final generation results through a cyclic process of noise addition and denoising:
\begin{equation}
    \hat{x}_{0}=\epsilon_{t}^{\theta}\left(x_{t}|c\right),\\
    x_{t-1}=\frac{1-\alpha_{t-1}+\sqrt{\alpha_{t-1}} \hat{x}_{0}}{1-\alpha_{t}\hat{x}_{0}}+\sigma_{t} z_{t},
\end{equation}
where $c$ is the conditions.

\subsection{Implicit Joint Distribution Matching}
Although the conditional DDPM-based methods can generate satisfactory results, the 1000-step denoising process renders these impressive outcomes inapplicable to real-time tasks in practice. If we aim to achieve fast sampling by reducing the denoising steps to a few dozen, the Gaussian assumption followed by DDPM's $p_{\theta}(x_{t-1}|x_{t})$ process (noise sampling from a unimodal distribution with a small value at each step) would no longer hold, and L2 reconstruction could not be applied to model the KL divergence. 

To address this issue, we introduce a GAN structure based on conditional denoising diffusion, which inherits the Conditional Gesture Generator and Conditional Gesture Discriminator, as shown in Figure~\ref{fig:arc}. We consider the time step t, audio control signal c, and seed gesture in conditional control simultaneously to implement a divide-and-conquer approach to effectively help the model learn the conditional denoising distribution. Additionally, we propose an adversarial learning strategy, using a conditional GAN to match the conditional distribution between $q(x_{t-1}|x_{t})$ and $p_{\theta}(x_{t-1}|x_{t})$, and adding random large noise between adjacent diffusion steps to achieve rapid denoising. The formulation can be summarized as follows:
\begin{equation}
\begin{aligned}
    \underset{\theta}{\min} \underset{D_{adv}}{\max} \sum\limits_{t>0} \mathbb{E}_{q(x_{t})}D_{adv}(q(x_{t-1}|x_{t})||p_{\theta}(x_{t-1}|x_{t})),
\end{aligned}
\end{equation}

More specifically, we will denote the Conditional Gesture Discriminator relying on time steps and control signals as $D_{\emptyset}(x_{t-1},x_{t},c,t)$. Here, $x_{t-1}$ and $x_{t}$ represent the noise at time steps $t-1$ and $t$, respectively, while $c$ denotes the joint of audio control signals and seed gestures. The conditional GAN's structure is shown in Figure 3. In our adversarial learning strategy, fake samples from distribution $p_{\theta}(x_{t-1}|x_{t})$ will be pitted against real samples from distribution $q(x_{t-1}|x_{t})$. Given the joint distribution formula $q(x_{t},x_{t-1})=\int dx_{0}q(x_{0})q(x_{t-1}|x_{0})q(x_{t}|x_{t-1})$, our discriminator can be trained as follows:
\begin{equation}
\begin{aligned}
    \underset{\theta}{\min} \sum\limits_{t>0} \mathbb{E}_{q(x_{0})q(x_{t-1}|x_{0})q(x_{t}|x_{t-1})} [-log(D_{\emptyset}(x_{t-1},x_{t},c,t))]\\
    +\mathbb{E}_{p_{\theta}(x_{t-1}|x_{t})}[-log(1-D_{\emptyset}(x_{t-1},x_{t},c,t))],
\end{aligned}
\end{equation}

This rewritten formulation (D-real) of conditional GAN allows us to model $p_{\theta}(x_{t-1}|x_{t})$ more flexibly. Given the training goal of the Conditional Gesture Discriminator, we also need to train the Conditional Gesture Generator, whose formula (G-fake) is defined as follows:
\begin{equation}
\begin{aligned}
    \underset{\theta}{\max} \mathbb{E}_{t~[1,T],q(x_{t})}\mathbb{E}_{p_{\theta}(x_{t-1}|x_{t})}[-log(-D_{\emptyset}(x_{t-1},x_{t},c,t))] ,
\end{aligned}
\end{equation}

\subsection{Explicit Geometric Constraints}
Although the conditional GAN based on control signals seems to reasonably address the large stride and small step count of high-speed sampling, the purely implicit adversarial learning for the connected $x_{t-1}$ and $x_{t}$ is statistically inefficient, particularly when $x_{t}$ is a high-dimensional redundant representation (which is often the case in human body sequences). This might also be the reason for the sharp decline in the performance of DDGAN on complex data distributions. Furthermore, we observe that after the adversarial process of the conditional GAN, the generator often produces gestures with artifacts. We speculate that this is because the purely implicit matching constraint on the joint distribution cannot provide geometric constraints for the gestures. Therefore, we propose an explicit geometric constraint for training the generator to improve the quality of the generated gestures. More specifically, following the approach of MDM~\cite{tevet2022human}, we directly predict clean gesture sequences with $\hat{x}_{0}=G(x_{t},z,c,t)$, and we employ a Huber loss~\cite{huber1992robust} to constrain its reconstruction loss:
\begin{equation}
\begin{aligned}
    \mathcal{L}_{recon}&=E_{x_{0}~q(x_{0}|c),t~[1,T]}[HuberLoss(x_{0}-\hat{x_{0}})],\\
\end{aligned}
\end{equation}
Lastly, we combine the implicit joint distribution matching constraint with the explicit gesture geometric constraint to train the Conditional Gesture Generator (also referred to as the denoiser).

\begin{table*}
	\centering
	\begin{tabular}{lccccccccc}
        \toprule[1.5pt]
		\multirow{2}{*}{Method} & \multicolumn{4}{c}{BEAT}   & \multicolumn{4}{c}{ZeroEGGs} \\
		& FGD$\downarrow$ & BA$\downarrow$ & DIV$\downarrow$ & $ms/frame$ & FGD$\downarrow$ & BA$\downarrow$ & DIV$\downarrow$ & $ms/frame$  \\ \hline
		GroundTruth & - & 0.93 & 0.88 & -& - &0.95&0.81 &- \\
            CAMN~\cite{liu2022beat} &258.4 &0.73 &0.61 &- &137.4 &0.69 &0.70\\
            Trimodal(re-train)~\cite{yoon2020speech} & 222.3 & 0.77 & 0.68 & - & 183.1 & 0.72 &0.69 &-\\
            \hline
		FreeTalker~\cite{yang2024freetalker} &147.2 & 0.84 & 0.53 &55.0 &82.7 &0.79 &0.63 &49.7\\
		DiffStyleGesture(re-train)~\cite{yang2023diffusestylegesture} & 100.3 & \textbf{0.89} & 0.71 & 75.0 & 78.4 & 0.82 &0.80 &77.4\\
		 DiffGes(re-train)~\cite{zhu2023taming} & 247.7 &0.66 &\textbf{0.82} &42.3 &157.4 &0.68 &0.88 &40.0\\\hline
		Ours & \textbf{84.0} & 0.87 & 0.76 &\textbf{3.2} & \textbf{54.7} & \textbf{0.83} & \textbf{0.88} &\textbf{4.1}\\ 
        \bottomrule[1.5pt]
	\end{tabular}
	\caption{\textbf{Comparison.} Bold indicates the best-performing method for each evaluation criterion. All experiments are conducted and evaluated on the BEAT and ZeroEgg dataset. The results for CAMN~\cite{liu2022beat} are reproduced using the official code, while the results for DiffStyleGesture~\cite{yang2023diffusestylegesture}, DiffGes~\cite{zhu2023taming} and Trimodal~\cite{yoon2020speech} are obtained by retraining the official code. Dashes ("–") indicate that specific metrics are not available. }
	\label{tab:3dpw}
\end{table*}
\begin{figure}
  \begin{center}
    \includegraphics[width=1\linewidth]{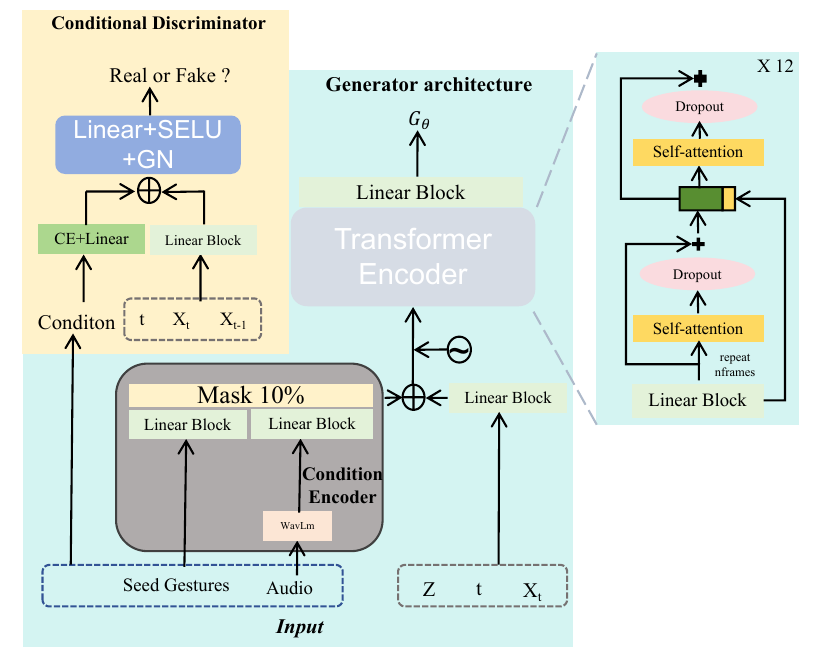}
  \end{center}
  \vspace{-5pt}
  \caption{\textbf{Conditional GAN architecture.} The Conditional Gesture Generator is employed for the audio-to-gesture task, where z is the latent variable sampled from a Gaussian distribution. It consists of a Transformer Encoder, Linear, and Mask components. The Conditional Gesture Discriminator is utilized to distinguish between genuine and counterfeit gesture sequences, comprising Linear, SELU, and Group Norm modules.}
  \vspace{-10pt}
  \label{fig:arc1}
  \end{figure}
  
\section{Experiments}
In this section, we will delve into the technical complexity of our model, comparisons with other frameworks, potential applications, and the validation of design choices for each module through ablation studies.\\
\textbf{Implementation Details. }As depicted in the right portion of Figure~\ref{fig:arc1}, we construct a Conditional Gesture Generator based on the Transformer architecture, comprising a 12-layer, 8-head encoder with default skip connections. The conditional control signal encoder consists of linear layers and WavLM. The linear block maps the seed pose and style label linearly. The audio input is initially encoded by WavLM to obtain spectral information, then linearly mapped to a space sharing the same dimensions as the former. As shown in the top-left corner of Figure~\ref{fig:arc1}, the Conditional Gesture Discriminator is a 7-layer MLP network, composed of linear layers, SELU activations, and GroupNorm layers. All models are trained employing the AdamW optimizer with a fixed learning rate of 1e-5. During training, we apply EMA decay to the optimizer. In the diffusion training phase, our mini-batch size is set at 64. Based on empirical values for GAN network training, we establish the generator's learning rate at 3e-5 and the discriminator's learning rate at 1.25e-4 for stable adversarial training. For the ZeroEGGs dataset training task, conditioning a denoising diffusion GAN requires approximately 80 hours on a single A100 GPU. During inference evaluation, we execute all experiments five times on a single V100 GPU and calculate the average.
\subsection{Datasets and Evaluation Metrics} 
\textbf{Datasets. }Our audio data is derived from the large-scale, high-quality \textbf{BEAT} dataset~\cite{liu2022beat} and the \textbf{ZeroEGGs}~\cite{ghorbani2023zeroeggs} dataset. The BEAT dataset encompasses 76 hours of multimodal speech data from 30 speakers, featuring 8 emotions and 4 languages. We selected the \textbf{English} data from demonstrations and conversations for training. On the other hand, the ZeroEGGs dataset is a highly expressive collection of high-quality gesture sequences, containing nearly 2 hours of full-body motion capture and monologue videos performed by a female actor in 19 distinct styles.

\textbf{Evaluation Metrics.} We concurrently assess the quality (FGD), beat alignment (BA), and diversity (DIV) of gestures generated by our method. Gesture quality is evaluated using Fréchet Gesture Distance (FGD)\cite{yoon2020speech}, which computes the distance between the latent feature distributions of generated and real gestures, thereby gauging gesture quality. Lower FGD values signify higher motion quality. Beat Alignment (BA)\cite{li2021ai} represents the inverse angular distance between audio and gesture beats, and is employed to appraise the similarity of gesture audio beats; the higher the value, the better the alignment with audio beats. DIV measures the L1 distance among multiple body gestures produced under the same control signal, with larger values indicating greater diversity. We adhere to the same evaluation implementation as in the BEAT paper.


\subsection{Comparison to Existing Methods}
\begin{figure}
  \begin{center}
    \includegraphics[width=1\linewidth]{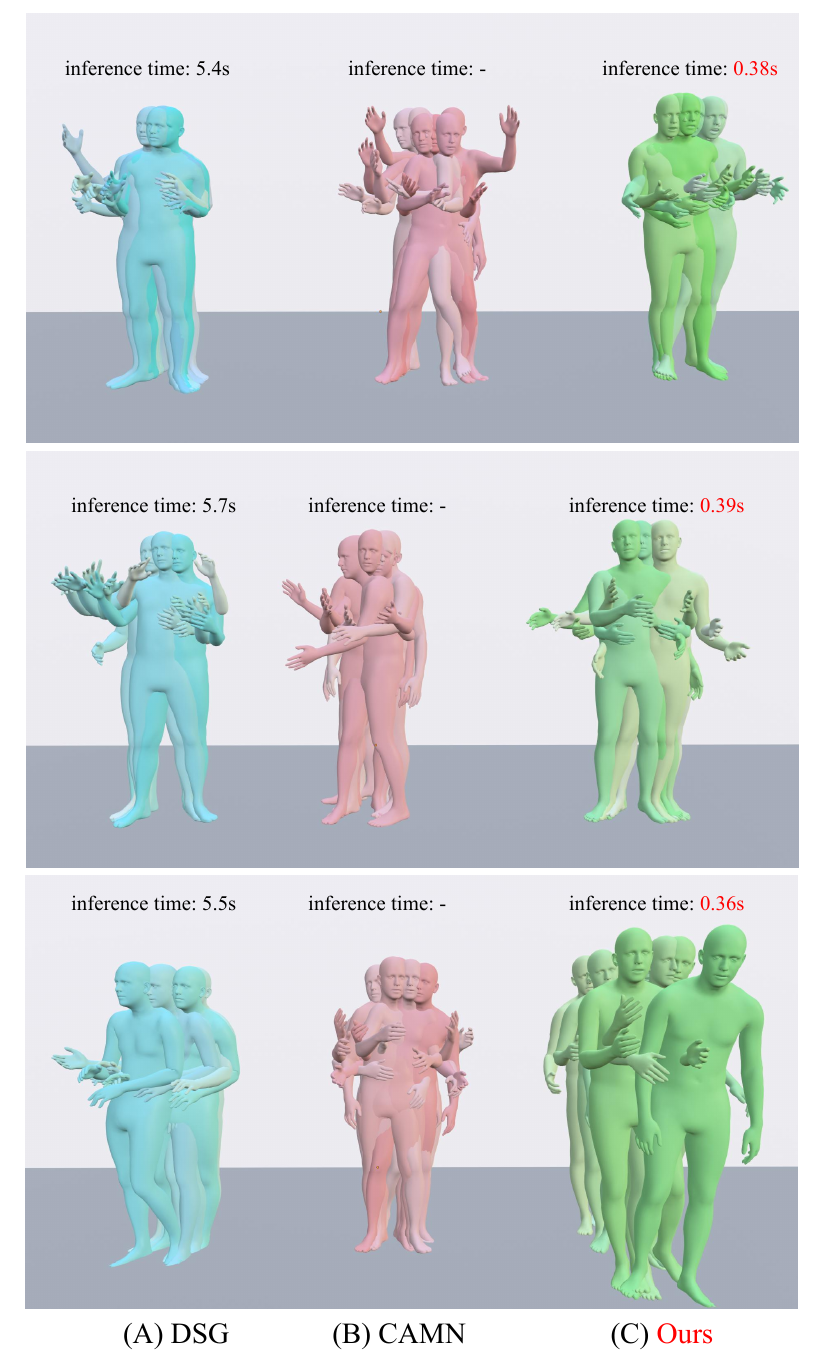}
  \end{center}
  \vspace{-10pt}
  \caption{In comparison to contemporary diffusion-based methods such as DSG and non-diffusion methods like CAMN, our approach achieves high-speed generation while ensuring the utmost generation quality.}
  \vspace{-20pt}
  \label{fig:qual}
  \end{figure}
We juxtapose our approach with existing diffusion-based and non-diffusion-based methods. On the BEAT dataset, as illustrated in Table~\ref{tab:3dpw}, non-diffusion methods frequently grapple with jitter issues in generated results, resulting in highly unnatural movements and inferior FGD scores. Conversely, diffusion-based methods are encumbered by the high generation time costs; for instance, DiffStyleGesture attains respectable FGD scores, but its generation speed approaches 75 milliseconds per frame, constraining its potential for deployment in real-time tasks. In contrast, our large-step, fewer-step method achieves a breakthrough in this domain, enhancing generation speed by over seven times to a mere 4.2 milliseconds per frame. This improvement allows a 4-second slice to be generated within 0.3 seconds, rendering it highly suitable for real-time tasks without any compromise in generation quality compared to DSG. 

Qualitative results, as displayed in Figure~\ref{fig:qual}, indicate that our method outperforms the previously best diffusion-based generation method, DSG, in terms of gesture effects while maintaining high-speed generation capabilities. The generation speed during the inference phase has improved nearly 13 times compared to DSG, necessitating only about 0.4 seconds to generate 80 frames (4 seconds) of gesture outcomes. This advancement renders our method suitable for real-time tasks, such as virtual human communication and real-time human-machine interaction. In contrast to non-diffusion methods like CAMN, our approach showcases more stable joint movements and avoids unnatural global jitter issues.

\begin{table*}
	\centering
	\begin{tabular}{lccccccccc}
        \toprule[1.5pt]
		\multirow{2}{*}{Method} & \multicolumn{4}{c}{BEAT}   & \multicolumn{4}{c}{ZeroEGGs} \\
		& FGD$\downarrow$ & BA$\uparrow$ & DIV$\uparrow$ & $ms/frame$ & FGD$\downarrow$ & BA$\uparrow$ & DIV$\uparrow$ & $ms/frame$$\downarrow$  \\ \hline
		DPM-Solver-1(DDIM) & 151.2 & 0.74 & 0.63 & 3.6& 99.8 &0.76&0.70 &4.5 \\
            DPM-Solver-2 & 144.7 & 0.70 & 0.68 & 5.8& 119.6 &0.72&0.67 &6.9 \\ \hline
            DiffGes+DDGAN & 374.8 & 0.53 & - & 3.5& 352.8 &0.55&- &\textbf{3.9} \\\hline
		Ours & \textbf{84.0} & 0.87 & 0.76 &\textbf{3.2} & \textbf{54.7} & \textbf{0.83} & \textbf{0.88} &4.1\\ 
        \bottomrule[1.5pt]
	\end{tabular}
	\caption{Comparative results with contemporary accelerated diffusion methods are presented. Ensuring fairness, all methods employ 10-step denoising. The original DDGAN, lacking geometric constraints, produces near-static outcomes, while DPM-Solver's higher-order components necessitate multiple denoising function calls, impacting generation speed. Our approach strikes a delicate balance between speed and quality. }
	\label{tab:3dpw}
\end{table*}

\subsection{Contrasting with Other Speedup Approaches}
In the table, we juxtapose our approach with other acceleration strategies tailored for diffusion-based generation methods. Specifically, our experimental results are contrasted with strategies employing DPM-Solver and the original DDGAN method. The experiments reveal that for the DPM-Solver strategy, its first-order Taylor expansion form corresponds to the well-known DDIM sampling strategy. Accelerating sampling by merely reducing the sampling step size often results in an imprecise approximation of complex multimodal distributions, causing a drastic deterioration in quality. The second-order Taylor expansion, due to the presence of the second-order derivative, necessitates invoking the denoising function twice at the midpoint position during sampling, consequently diminishing generation speed while providing limited enhancement in generation quality. Higher-order Taylor expansions demand even more frequent calls to the denoising function, which contradicts the primary objective of acceleration.

Furthermore, we directly integrate the recent work, DDGAN, with DiffGes for comparison. Specifically, we train an unconditional discriminator for DiffGes on the BEAT dataset, eliminate explicit geometric loss, and compare it with our method. The results, as displayed in the table, indicate that the DiffGes+DDGAN configuration performs exceedingly poorly concerning the generated global gesture quality. This is attributable to the fact that, unlike images, human representations often possess more stringent geometric conditions and necessitate more specific constraints.

\subsection{Ablation Study}
\begin{table}
\centering
\begin{tabular}{lccccccccc}
\toprule[1.5pt]
\multirow{1}{*}{Method} &\multicolumn{1}{c}{$SA$}$\uparrow$ &\multicolumn{1}{c}{$HL$} $\uparrow$&\multicolumn{1}{c}{$GA$}$\uparrow$ &\multicolumn{1}{c}{$ms/frame$}$\downarrow$ \\ \hline
Ground Truth &0.86  &4.41 &4.72 &-\\ 
DiffStyleGesture &0.80  &3.97 &3.74 &75.0\\ \hline
Ours &0.76 &4.03 &3.82 &3.2\\
w/o WavLM &0.67 &3.79 &3.77 &4.01 \\
w/ GRU &0.72 & 3.71 &3.76 & 3.08\\
\hline
\end{tabular}
\vspace{-10pt}
\caption{\textbf{Ablation study on structure.} 'w/' means with and 'w/o' means without. }
\label{tab:abl1}
\end{table}
In this section, we verify the efficacy of our designed key modules. All experiments are carried out on the BEAT dataset. Given that the BEAT dataset encompasses data from 30 English speakers, resulting in substantial training costs, we opt for sequences from three speakers for validation purposes. All modules employ identical experimental configurations to guarantee fairness.\\
\textbf{Structural Ablation.} As illustrated in Table~\ref{tab:abl1}, concerning audio control information, when substituting WavLM features with the first 13 coefficients of traditional MFCC, the generated results' performance fails to meet users' subjective perception due to the absence of further extraction of semantic and emotional information from audio input latencies. As a result, the scores markedly decline. Upon replacing our Transformer-Encoder with a GRU structure, user scores experience a significant reduction. This is attributable to the inherent asynchrony between audio and gestures, where the attention mechanism proves more reliable in capturing such asynchronous relationships compared to GRU. Simultaneously, given that in actual communication, arms generally express emotional styles while fingers convey semantics, we devise a decoupled structure for both, employing learnable weights for balance. This approach allows our method to surpass DSG. It is worth noting that during structural ablation, we consistently utilize the model comprising 20 diffusion steps. \\
For the ablation experiment evaluation, we primarily assess human subjective perception consistency metrics, evaluating from two dimensions. The first two dimensions follow GENEA~\cite{yoon2022genea}, assessing human likeness (HL) and gesture-audio appropriateness (GA). The third dimension is gesture style matching (SA). To understand our method's true visual performance, we conducted user studies comparing generated gesture sequences with real motion capture data. Evaluation clips range from 16 to 60 seconds, averaging 38.1 seconds. We recruited 60 participants in three groups, with scores ranging from 5 to 1, labeled as "excellent," "good," "average," "below average," and "poor."\\
\begin{table}
\centering
\begin{tabular}{lccccccccc}
\toprule[1.5pt]
\multirow{1}{*}{Steps} &\multicolumn{1}{c}{$FGD$}$\downarrow$ &\multicolumn{1}{c}{$BA$}$\uparrow$ &\multicolumn{1}{c}{$SA$}$\uparrow$ &\multicolumn{1}{c}{$ms/frame$}$\downarrow$ \\ \hline
1  &1174.4  &0.61 &- &2.2\\ 
5  &298.17  &0.74 &0.61 &2.8\\ 
10  &84.0  &\textbf{0.87} &\textbf{0.76} &3.6\\ 
20  &84.6  &0.87 &0.71 &5.1\\ 
30  &83.7  &0.85 &0.63 &7.7\\ 
50  &\textbf{83.5}  &0.86 &0.65 &9.3\\ 
\hline
\end{tabular}
\vspace{-10pt}
\caption{Ablation study on sampling steps.}
\label{tab:abl2}
\end{table}
\textbf{Sampling Step Ablation.} In this experiment, we studied the impact of different sampling steps on model performance on the BEAT dataset. Specifically, we trained models with the same structure using 1, 5, 10, 20, 30, and 50 steps, respectively. The final results, as shown in Table~\ref{tab:abl2}, indicate that fewer steps result in a higher sampling speed. However, an excessively large step size causes a sharp decline in quality. When the number of steps is 1, our structure degenerates into a traditional GAN model. From the results, we observe that the improvement in FGD becomes more gradual after 10 steps, but an increase in the number of steps also leads to slower speed. Therefore, we compromise and choose the result with 20 steps as the final demo model. \\
\textbf{Influence of Geometric Constraints.} In this experiment, we validate the role of the explicit gesture geometric constraints mentioned in Section 3.3 on the BEAT dataset. When the geometric constraint weight is set to 0, our method degenerates into a purely implicit joint distribution matching constraint for $x_{t}$ and $x_{t-1}$, which is a statistically inefficient and original DDGAN incapable of handling high-dimensional feature representations, resulting in a significant decline in gesture quality, as shown in Table~\ref{tab:abl3}. With the addition of explicit geometric constraints, the gesture quality is noticeably improved. Based on our experience, we set the weights to 1, 10, and 100 for the experiments and found that the results did not differ significantly. In this paper, we use the model with a weight of 10 as the final demo. 
\begin{table}
\centering
\begin{tabular}{lccccccccc}
\toprule[1.5pt]
\multirow{1}{*}{Weights} &\multicolumn{1}{c}{$FGD$}$\downarrow$ &\multicolumn{1}{c}{$BA$}$\uparrow$ &\multicolumn{1}{c}{$SA$}$\uparrow$\\ \hline
0  &988.7  &0.52 &-\\ 
1  &88.2  &0.85 &-\\ 
10  &\textbf{84.6}  &\textbf{0.87} &0.71\\ 
100  &86.7  &0.83 &\textbf{0.74}\\ 
\hline
\end{tabular}
\vspace{-10pt}
\caption{Ablation study on Geometric Constraints. All models adopt 20 diffusion steps. }
\vspace{-10pt}
\label{tab:abl3}
\end{table}


\section{Conclusions and Discussions}
In conclusion, this paper has presented an innovative approach to address the limitations of diffusion models in the generation of co-speech gestures. The focus has been on enhancing the generation speed and modeling complex distributions between multiple sampling steps. Unlike the acceleration strategies of previous diffusion-based text-to-motion methodologies, such as those utilized in DDIM, our method avoids the stacking of small-step noise. Instead, we have introduced a conditional GAN to model the complex denoising distribution under audio control signals, enabling our model to denoise with larger and fewer steps, thus improving efficiency.

Furthermore, our approach addresses the statistical inefficiency of DDGAN in image generation and its inability to accommodate high-dimensional feature representations. By moving away from the original purely implicit matching constraint on joint distribution and incorporating explicit gesture geometric constraints, our method achieves significant acceleration while preserving high-fidelity generation results. This approach provides a promising direction for future real-time gesture generation tasks. We pledge to make our code repository publicly available, allowing future research to focus on improving the proposed method~\cite{han2024reindiffuse}, exploring its applicability in other domains (such as providing more diverse generated data for motion capture~\cite{cheng2023bopr, liang2024ropetp} and whole body generations~\cite{yu2023signavatars}), and investigating the integration of additional constraints to enhance the quality and realism of generated gestures. In addition, some target detection methods~\cite{sun2024bidirectional} are used to extract details of hand and facial information to optimize the generated results.


{\small
\bibliographystyle{ieee_fullname}
\bibliography{egbib}
}

\end{document}